\documentclass[10pt,conference,a4paper]{IEEEtran}
\usepackage{cite}
\usepackage{graphicx}
\usepackage[cmex10]{amsmath}
\usepackage{url}
\usepackage{algorithm}
\usepackage{amssymb}
\usepackage{tabularx}
\usepackage{float}
\usepackage{fancybox}
\usepackage{longtable}
\usepackage{verbatim}
\usepackage{flushend}
\usepackage{etoolbox}

\makeatletter
\patchcmd{\@makecaption}
{\scshape}
{}
{}
{}
\makeatletter
\patchcmd{\@makecaption}
{\\}
{.\ }
{}
{}
\makeatother
\def\tablename{Table}

\begin{document}
\sloppy

\IEEEoverridecommandlockouts

\graphicspath{{}}
%
\title{On the Application of Network Slicing for 5G-V2X}

\author{\begin{minipage}[t]{1.0\linewidth}\centering Hamza Khan$^\star$ Petri Luoto$^\dag$, Mehdi Bennis$^\star$, Matti Latva-aho$^\star$\end{minipage}\hspace{-1mm} \\

\begin{minipage}[t]{0.5\linewidth}\centering $^\star$Centre for Wireless Communications\\University of Oulu, Finland\\P.O. Box 4500, FI-90014 Oulu \{hamza.khan, mehdi.bennis, matti.latva-aho\}@oulu.fi\end{minipage}

\begin{minipage}[t]{0.33\linewidth}\centering $^\dag$MediaTek Wireless Finland Oy\\Elektroniikkatie 16, \\ FI-90590 Oulu\\ petri.luoto@mediatek.com\end{minipage}

}


%


\maketitle

\begin{abstract}
Ultra-reliable vehicle-to-everything (V2X) communication is essential for enabling the next generation of intelligent vehicles. V2X communication refers to the exchange of information between vehicle and infrastructure (V2I) or between vehicles (V2V).  Network slicing is one of the promising technologies for the next generation of connected devices, creating several logical networks on a common and programmable physical infrastructure. Following this idea, we propose a network slicing based communication model for vehicular networks. In this paper, we have modelled a multi-lane highway scenario with vehicles having heterogeneous traffic requirements. Autonomous driving slice (exchanges safety messages) and infotainment slice (provides video stream) are the two logical slices created on a common infrastructure. In addition, a relaying approach is utilized to improve the performance of low signal-to-interference-plus-noise-ratio (SINR) video streaming vehicles. These low SINR vehicles are served by other infotainment vehicles, which have high quality V2V and V2I link and are not serving as autonomous driving slice access point. An extensive Long Term Evolution Advanced (LTE-A) system level simulator is used to evaluate the performance of the proposed method, in which it is shown that the proposed network slicing approach increases the packet reception ratio (PRR) from 31.15\% to 99.47\%.
\end{abstract}


%
\IEEEpeerreviewmaketitle

\section{Introduction}\label{sec:intro}
Advancement in information technology has led the automotive sector to pursue the dream of intelligent vehicles. Automatic breaking system, lane departure warning system and cruise control are various forms of intelligence that are present in today's vehicles. An autonomous vehicle is the ultimate realization of intelligent vehicles for which ultra-reliable vehicle-to-everything (V2X) communication is a key enabler \cite{ultra}. Autonomous vehicles aim at reducing a multitude of problems in the current transportation system.

A significant proportion of fatalities occur at intersections and roundabouts where the communication between vehicles are needed. 
Research shows that human error is completely or partially the cause of accident in 90\% of the cases \cite{national}. V2X communication has been researched for more than a decade because of its significance in safe transportation. 

In early 2000's IEEE 802.11p based technology dedicated short range communication (DSRC) appeared as the key enabler for V2X communication \cite{V2X_IEEE}. However, DSRC suffers from unbounded latency and cannot guarantee the quality-of-service (QoS) as shown in \cite{unbounded}. Currently, the focus has shifted towards cellular V2X communication primarily because the infrastructure is already deployed which enables low-latency and ultra-reliable V2X communication \cite{V2X_LTE2}.

According to the European Telecommunications Standards Institute (ETSI), the end-to-end latency for vehicular communication is 100 ms for a message size of about 1600 bytes with transmission reliability of 99.999\% \cite{speed}. These safety messages carry information i.e. vehicle location, kinematics, cooperative cruise control, collision warning, etc \cite{CAM}.  

Radio resource allocation plays an important role in achieving the QoS requirement of vehicular communication \cite{RRM}. Improvement in road safety, traffic efficiency and provisioning of local services are the core design considerations for V2X communication. To address these requirements, authors in \cite{claudia} propose the design of customized network slices for V2X services.  
Performance of cellular V2X communication is analysed in \cite{Petri1} and a relaying mechanism to serve low SINR cell edge vehicles is proposed in \cite{Petri2}. Analysis of multicast and unicast transmission for LTE-V2X network is done in \cite{multi-unicast}. Decentralized resource allocation approach for reliable communication was proposed in \cite{RRM2-Ikram}.

\begin{figure}
	\centering
	\includegraphics[trim = 0mm 0mm 0mm 0mm, clip, width=0.55\textwidth]{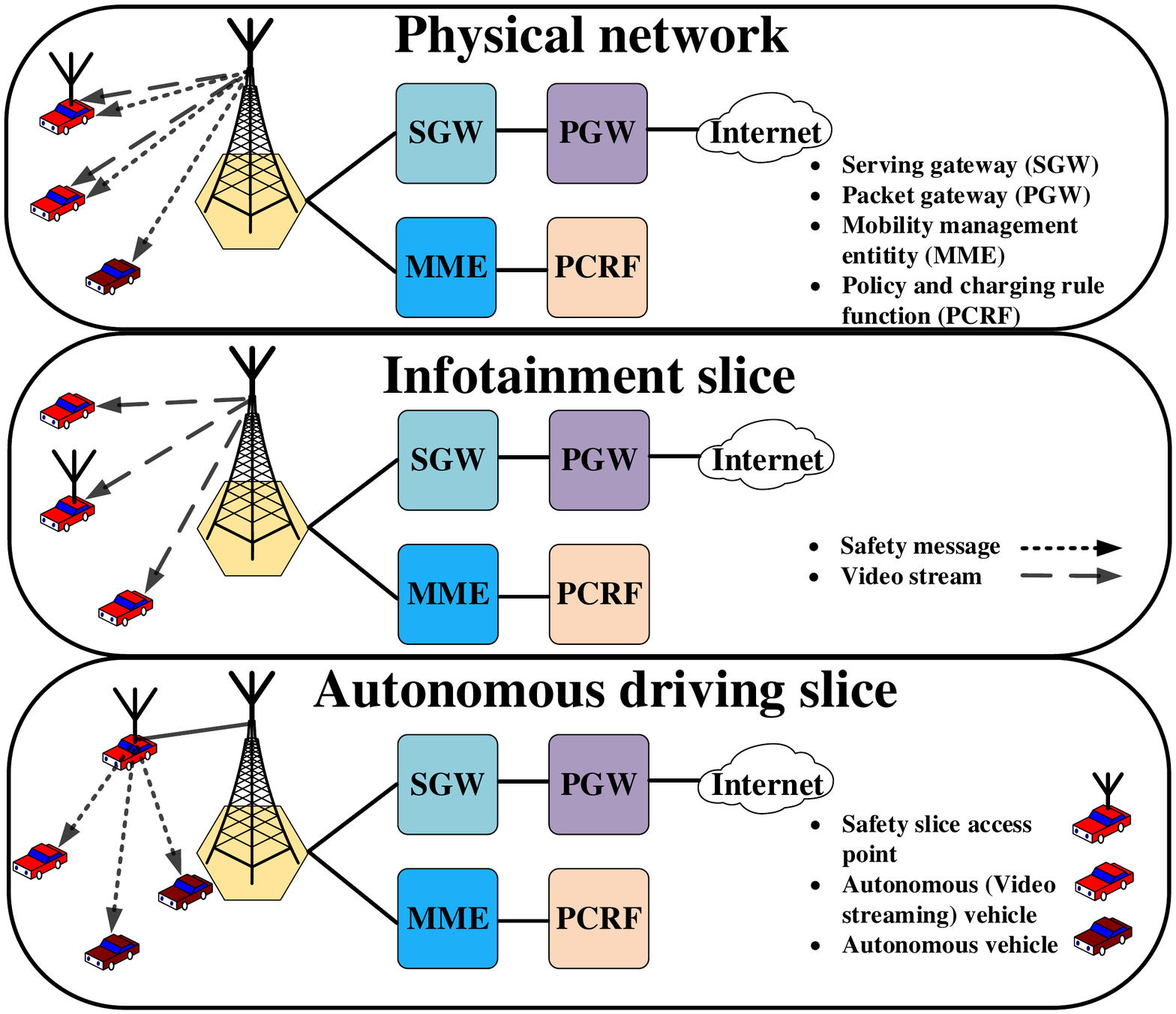}
	\caption{\label{NS}Illustration of network slicing for V2X.}
\end{figure}

In this paper we introduce a network slicing based V2X communication model in the downlink direction with heterogeneous traffic. Network slicing is defined as the concept of creating multiple logical networks on a shared physical infrastructure \cite{slice1}. The idea of network slicing for V2X communication was proposed in \cite{metis_slicing}. Fig. \ref{NS} illustrates the network slicing model used in this work. Autonomous driving slice and infotainment slice are the two sub-slices created on a common physical infrastructure to provide isolated services.  The former is responsible for the reliable exchange of safety messages between vehicles while the latter provides video streaming services. Heterogeneous traffic consists of safety messages and video streaming service.  

Road side unit (RSU) is responsible for providing every type of service in the legacy LTE-A network. In our proposed network slicing method, LTE-A RSU provides infotainment services only and decides which video streaming vehicle will serve as autonomous driving slice access point based on V2I and V2V link quality. 
These access points function as virtual RSUs, which are responsible for the reliable exchange of safety messages between vehicles of autonomous driving slice. Different vehicular densities are considered in this work. Simulation guidelines for the simulator are recommended by the International Telecommunication Union (ITU) \cite{WINII}.

This paper is structured as follows. Section II explains the system and link model. Network slicing for V2X communication is explained in Section III. Performance evaluation of the proposed technique is provided in Section IV. Concluding remarks are provided in Section V.

\section{System Model} \label{sec:model_system}
Single-input multiple-output (SIMO) transmission with orthogonal frequency-division multiple access (OFDMA) is considered. The network consists of a set of RSUs $R$, a set of vehicles $V$ (all vehicles require safety messages among which a subset of vehicle also requests infotainment services), a set $K \subset V$ of video streaming vehicles and a set of access points (video streaming vehicles which have high quality V2I and V2V links) $S \subset K$. RSU and access points have resource sets $L$ and $M$ which consist of physical resource blocks. Interference in the network occurs when resources are reused by multiple RSUs or access points. Each node in the network has $N_\text{r}$ receive antennas and $N_\text{t}$ transmit antennas. 

The communication channel from RSU $r$ to video streaming vehicle $k$ over subcarrier $l$ is denoted as $\textbf{h}_{r,k}^{l}(t)$. Similarly the channel vector from access point $s$ to vehicle $q$ is denoted as $\textbf{h}_{s,q}^{m}(t)$, where $Q\in V\setminus S$. The received signal by video streaming vehicle at time $t$ is given by (\ref{receivedsignalvideo}) and the received signal by autonomous vehicle is given by (\ref{receivedsignalsafety}):

\begin{equation}
\textbf{y}_{r,k}^{l}(t) = \textbf{h}_{r,k}^{l}(t)\textbf{x}_{r,k}^{l}(t)+\sum_{p\in R\setminus r}\textbf{h}_{p,k}^{l}(t)\textbf{x}_{p,k}^{l}(t)+\textbf{z}_{r,k}^{l}(t)
\label{receivedsignalvideo},
\end{equation}

\begin{equation}
\textbf{y}_{s,q}^{m}(t) = \textbf{h}_{s,q}^{m}(t)\textbf{x}_{s,q}^{m}(t)+\sum_{b\in S\setminus s}\textbf{h}_{b,q}^{m}(t)\textbf{x}_{b,q}^{m}(t)+\textbf{z}_{s,q}^{m}(t)
\label{receivedsignalsafety},
\end{equation}
where $\textbf{x}_{r,k}^{l}(t)$ is transmitted signal from the RSU $r$ to video streaming vehicle $k$,  $\textbf{h}_{p,k}^{l}(t)$ is the channel vector from interfering RSU to the \emph{k}th vehicle and $\textbf{z}_{r,k}^{l}(t)$ is the additive noise. Similarly, $\textbf{x}_{s,q}^{m}(t)$ is the transmitted signal from access point $s$ to autonomous vehicle $q$ and the interfering channel vector is $\textbf{h}_{b,q}^{m}(t)$.
The received SINR using maximum ratio combining for video streaming vehicle over subcarrier $l$ is given by

\begin{equation}
\textbf{SINR}_{r,k}^{l}(t) = \frac{|\textbf{h}_{r,k}^{l}(t)|^2}{|\sum_{p\in R\setminus r}\textbf{h}_{p,k}^{l}(t)|^2+\sigma_{z}^{2}}
\label{SINR},
\end{equation}
where $\sigma_{z}^{2}$ is the noise variance. Similar expression of SINR is achieved for autonomous vehicles. Inter-slice interference is mitigated using different carrier frequencies. Autonomous driving slice uses frequency of 2 GHz and infotainment slice uses frequency of 5.9 GHz. 

\begin{figure}
	\centering
	\vspace{-10pt}
	\includegraphics[trim = 0mm 0mm 0mm 0mm, clip, width=0.5\textwidth]{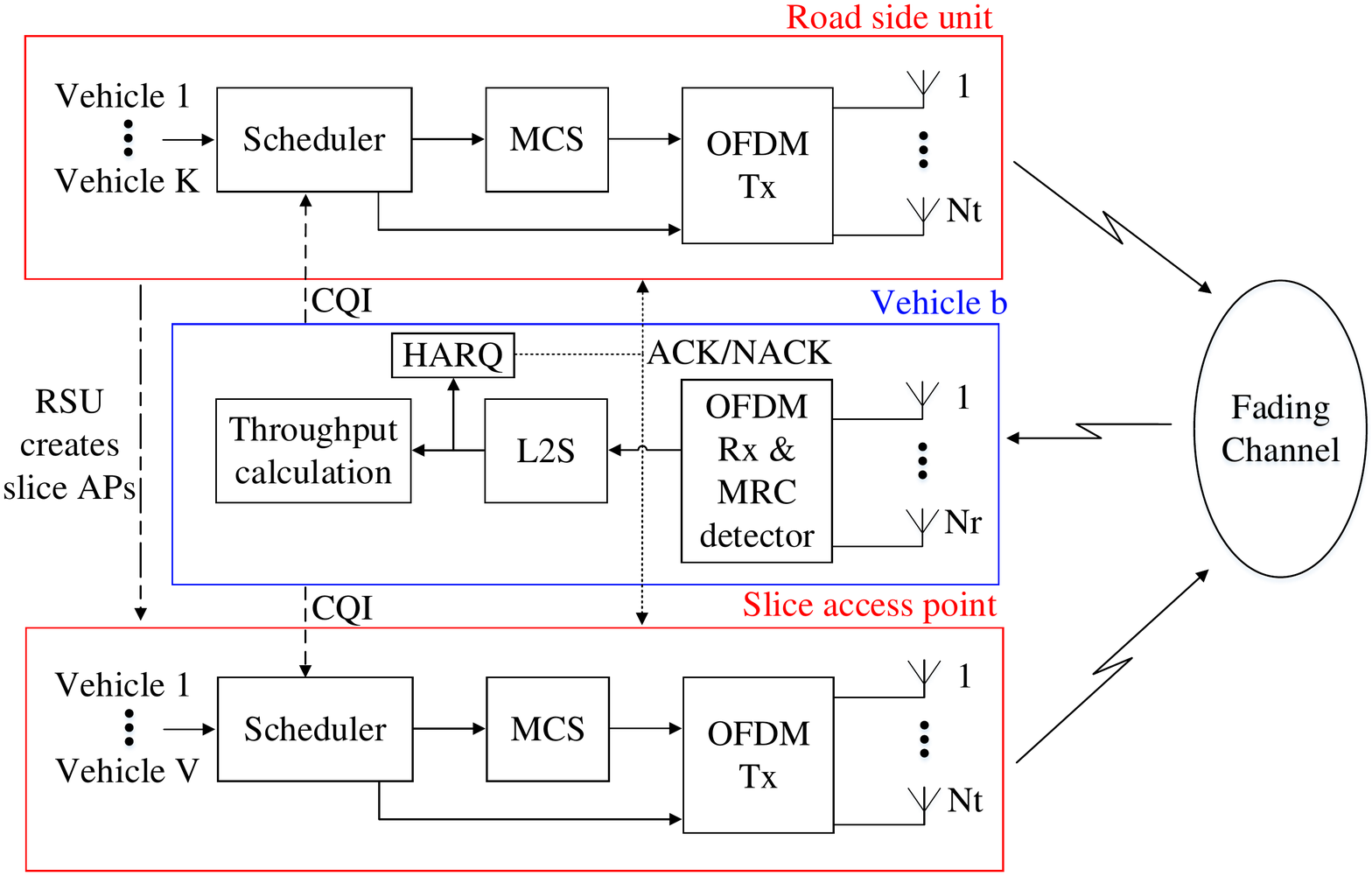}
	\vspace{-20pt}
	\caption{\label{Link_Model}Link model of V2X network with network slicing.}
\end{figure}

Fig. \ref{Link_Model} illustrates the link modelling of network slicing. Wireless link between the infotainment vehicles and RSU is modelled using geometry-based stochastic channel model (GSCM) given by \cite{WINII} and path loss model follows the macro to relay model. While, independent and identically distributed (i.i.d.) channel is assumed for autonomous driving vehicles and it follows the path loss model given in \cite{V2V_pathloss}.

Link-to-system (L2S) interface is used in the modelling of radio links for which channel quality information (CQI) is the most important parameter. CQI information is transmitted by each vehicle and it is used to determine the modulation and coding scheme (MCS). Network resources are utilized efficiently when proportional fair (PF) scheduling algorithm is employed in the RSU and slice access points. PF scheduler guarantees a minimum level of service for all vehicles while maximizing the throughput. RSU only schedules those vehicle which require infotainment service while, autonomous driving slice access point schedules all the vehicles. Effect of inter-symbol interference is not considered because the cyclic prefix is assumed to be longer than the delay spread. 

Maximal ratio combining (MRC) detector is used at the receiving node to exploit spatial diversity. SINR calculations are performed at the receiving nodes for every physical resource block (PRB). It is assumed that symbols are synchronized in both time and frequency domain. Mutual information based effective SINR mapping (MIESM) is used to reduce the computational overhead caused by the exact modelling of radio links. MIESM maps the SINR to corresponding mutual information curves.

Mutual information is then mapped to the frame error curves of corresponding MCS to evaluate erroneous transmissions. An acknowledgement (ACK) is generated for successful transmission and a negative acknowledgement (NACK) represents failed transmissions. Hybrid automatic repeat request (HARQ) is used for the re-transmission of failed frames.

\section{Network Slicing}\label{sec:slicing}

Network slicing is the process of slicing a physical network into multiple logical networks each of which can be optimized for a certain type of application. Network function virtualization (NFV) is an important technology, which serves as basis for network slicing. NFV makes it possible to reconfigure networks with software and it also virtualizes the network resources. In the current deployed network architecture each service has its own predefined service function chain. In network slicing NFV enabled nodes can provide these service functions.


Whenever a vehicle enters the network it requests RSU $r$ to provide a set of services. The main disadvantage of today's wireless network is that a number of services are provided with the same physical infrastructure (RSU) with no elasticity and it is referred as direct RSU communication in this work. 
In network slicing we create multiple slices depending upon the service requirement of vehicles and each service type has its designated service function chain. In this work an autonomous driving slice and an infotainment slice are created. Autonomous vehicles receive the safety messages from its slice access point, which in this work is a video streaming vehicle. For infotainment services RSU acts as an access point.

RSU measures the similarity between video streaming vehicle $K$ and autonomous driving vehicle $Q$ on the basis of geographical information. Building the similarity matrix $\textbf{C}$ is done such that vehicles which are close to each other have high similarity while vehicles far from each other have low similarity. 
Access points for new service function chain is created as per clustering algorithm of \cite{clustering_tutorial}. 
An important input to this algorithm is the number of video streaming vehicles (which can become autonomous driving slice access point). A video streaming user can only become autonomous driving slice access point if it has good quality V2I and V2V link. When the network is dense small number of slice access points can provide reliable connectivity but when the network become sparse i.e. vehicles are far from each other, then a large number of slice access points are required to provide high quality V2V links.

Distance based similarity matrix $\textbf{C}$ is formulated at the RSU, where $c_{k,q}$ corresponds to the ($k,q$)-th entry. Similarity $c$ between the video streaming user $k$ and autonomous driving vehicle $q$ is measured by the Gaussian similarity function given as \cite{Gaussian_similarity}:

\begin{equation}
\text{c}_{k,q} = \frac{-||\text{w}_{k} - \text{w}_{q}||}{2\sigma^{2}}
\label{receivedsignal},
\end{equation}
where $w_k$ is the location of the video streaming vehicle and $w_q$ is the location of the autonomous driving vehicle. Impact of neighbourhood size is controlled by $\sigma$. If the neighbourhood size is fixed, an appropriate choice of autonomous driving slice access points can be formulated as follows \cite{clustering_tutorial}:

\begin{equation}\label{similarity}
f = \arg\max_{e}(\zeta_{e+1}-\zeta_e), \hspace{0.2cm} e = 1,\ldots,\max|\mathcal{Q}|,
\end{equation}
where $\zeta_e$ is the $e$-th smallest eigenvalue. These eigenvalues are calculated from the network Laplacian matrix. RSU computes an unnormalized graph Laplacian matrix as the difference between diagonal matrix D and similarity matrix C, where the $k$-th diagonal entry of D is  $\sum_{q=1}^{\max|\mathcal{Q}|}c_{k,q}$. In the upcoming section a performance analysis of the vehicular network with different network densities is performed. 

\section{Performance Analysis}\label{sec:Analysis}
System level simulators developed under the guidelines of ITU provide accurate results on the expected performance. 
Simulation results provide a better understanding of key issues such as outage analysis, radio resource allocation and interference management. In this work we analyze the performance of network slicing based cellular V2X network with two sub-slices. Simulation setup in this work considers a six lane highway layout as proposed in \cite{Petri2}. Vehicles in three lanes are moving in the left direction and vehicles in the remaining three lanes are moving towards the right direction. Vehicles on the highway are moving with a speed of 140 km/h. The RSU network is situated along the highway at a distance of 35 m from the first lane. 

Autonomous vehicles are moving in each lane of the highway over a 10 km long stretch. To model different network densities inter-vehicular distance between vehicles is varied. Keeping the inter-vehicular distance small results in dense network and on the other hand large inter-vehicular distance results in sparser network. RSU repeats network slicing procedure after 100 ms to get updated information about V2I and V2V links. 
After updating link status RSU creates several safety access points based on clustering algorithm proposed in \cite{clustering_tutorial}. Each access points only serves those autonomous vehicles with which it can guarantee reliable communication.

Autonomous driving slice access points forward the safety messages to those autonomous vehicles which are allocated to it by the RSU. RSU forwards infotainment packets to video streaming vehicles. V2I communication in the downlink uses transmit power of 46 dBm for a message size of 1000 bits and a transmission periodicity of 1 ms. Similarly, transmission power for V2V communication is 20 dBm for a message size of 1600 bytes and periodicity of 100 ms. 
The rest of simulation parameters are summarized in Table \ref{params}. Performance of network slicing alone and network slicing with relaying is analysed in this work. The latter method was implemented to analyze throughput gain of the infotainment slice.

\begin{table}
	\centering
	\caption{Simulator parameters.}%
	\label{params}
	\begin{tabular}{|p{3cm}|p{4.2cm}|}
		\hline
		\textbf{Parameter} & \textbf{Assumption} \\ \hline
		Duplex mode & FDD \\ \hline
		System Tx bandwidth & 10 MHz \\ \hline
		& V2I: \hspace{2.5pt}2 GHz, \\
		\vspace{-10pt}Carrier frequency				&	V2V: 5.9 GHz \cite{3GPP_frequency}\\ \hline
		Antenna configuration & 1 Tx $\times$ 2 Rx\\\hline
		Receiver type & Maximum ratio combining \\ \hline
		Vehicle speed & 140 km/h \\ \hline
		& Scenario 1:  \hspace{8pt}1-100 m, \\
		Inter-vehicular Distance & Scenario 2: 100-200 m, \\
		& Scenario 3: 200-300 m \\ \hline	
		Scheduler & Proportional fair \\ \hline	
		& V2I: \hspace{2pt}46 dBm, \\
		\vspace{-10pt}	Transmission power &	V2V: 20 dBm \\ \hline
		L2S interface metric & MIESM \\ \hline
		& V2I: \hspace{0.1pt} 1000 bits/1 Hz, \\
		\vspace{-10pt}	Packet size &	V2V: 12800 bits/100 Hz \\ \hline
		Synchronization & Time and frequency synchronized \\ \hline
		HARQ & Chase combining \\ \hline
		Inter RSU distance & 1732 m\\ \hline
	\end{tabular}
\end{table}

\subsection{Network Slicing}\label{sec:NS} 
Performance of the proposed network slicing solution without relaying is analysed in this section. Description of the compared technologies are provided in Table \ref{description}. Direct RSU communication represents the legacy cellular network where every type of service is provided with the same infrastructure. Proposed method of \cite{Petri2} (direct RSU communication with relaying) provides a mechanism of relaying low SINR vehicles to achieve performance gains. Network slicing approach with and without relaying is proposed in this work.



\begin{figure}
	\centering
	\includegraphics[trim = 0mm 0mm 0mm 0mm, clip, width=0.5\textwidth]{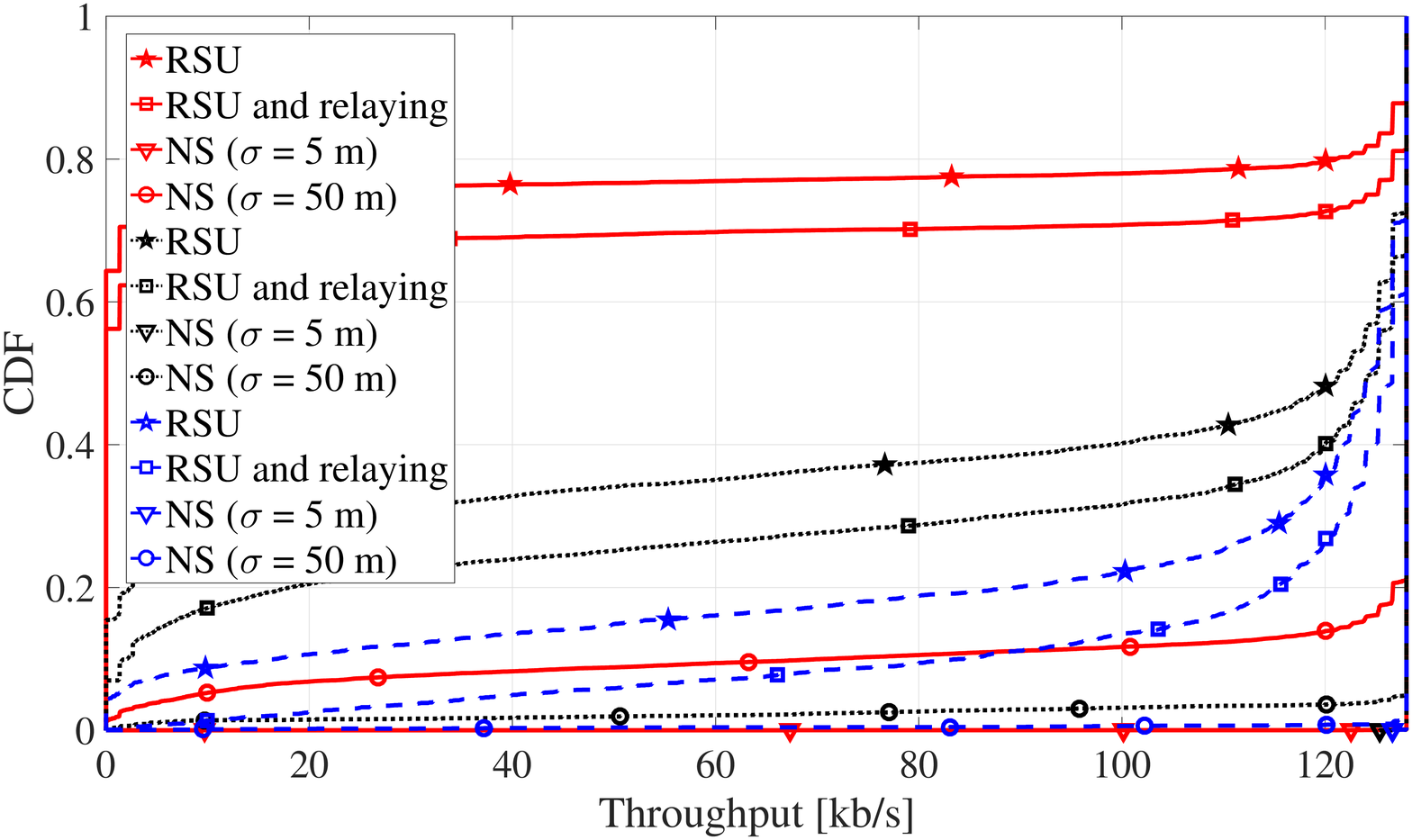}
	\caption{\label{Safety}CDF of the throughput of autonomous  slice for different technologies.}
\end{figure}

The cumulative distribution function (CDF) of the throughput of safety messages is shown in Fig. \ref{Safety}. 
Red color represents a dense network when the inter-vehicular distance is between 1-100 m, black color represents inter-vehicular distance of 100-200 m and blue color is used for sparse network with inter-vehicular distance 200-300 m. 

Network slicing procedure creates several slice access points on the basis of neighbourhood size. From Fig. \ref{Safety} it can be seen that when $\sigma = 50$ m the performance is lower compared to $\sigma = 5$ m. When $\sigma = 5$ m, there are $s = 115$ access points in the network for inter-vehicular distance 1-100 m. Inter access point interference is reduced due to orthogonal allocation of resource blocks to nearby transmitters.  

When $\sigma = 50$ m, there are $s = 13$ access points for the same inter-vehicular distance which increases the inter access point interference due to the increased number of vehicles under each access point. Resource blocks are reused more often with small number of access points and performance is degraded since PRB based interference model is used. Choice of neighbourhood size is effected by the density of network, large value of $\sigma$ in dense networks could lead to overloaded access points.

3GPP has standardized packet reception ratio (PRR) as a performance metric \cite{3GPP_frequency}. PRR for one transmitted packet is calculated as the ratio of number of vehicles which successfully receive the packet to the total number of vehicles which are present in the locality. Table \ref{PRR_safety} compares the PRR of the direct RSU communication and network slicing with different neighbourhood size. Dense networks with inter-vehicular distance of 1-100 m show huge improvements when network slicing is introduced. PRR of dense network with direct RSU communication is 31.15 \% which is increased to 99.47 \% when network slicing with neighbourhood size of 5 m is introduced. Sparse network with inter-vehicular distance of 200-300 m achieves the maximum PRR of 99.78 \% when network slicing is used.

\begin{table}
	\centering
	\caption{Description of compared technologies.}%
	\label{description}
	{
		\begin{tabular}{{|p{3cm}|p{4.2cm}|}}
			\hline
			\textbf{Technology} & \textbf{Description} \\
			\hline
			RSU & LTE-A RSU provides coverage for multiple service. \\ \hline
			RSU and relaying 
			& Coverage is provided by LTE-A RSU and vehicles with low SINR are offloaded to be served by other vehicles having high quality V2I and V2V link.  \cite{Petri2} \\ \hline
			Network slicing (NS) & LTE-A RSU provides video streaming services and autonomous slice access points exchange safety messages. \\ \hline
			NS and relaying & Service is provided by designated network slices and low SINR video streaming vehicles are relayed from other infotainment vehicles having high quality links. \\ \hline
			
	\end{tabular}}
\end{table}


\begin{figure}
	\centering
	\includegraphics[trim = 0mm 0mm 0mm 0mm, clip, width=0.5\textwidth]{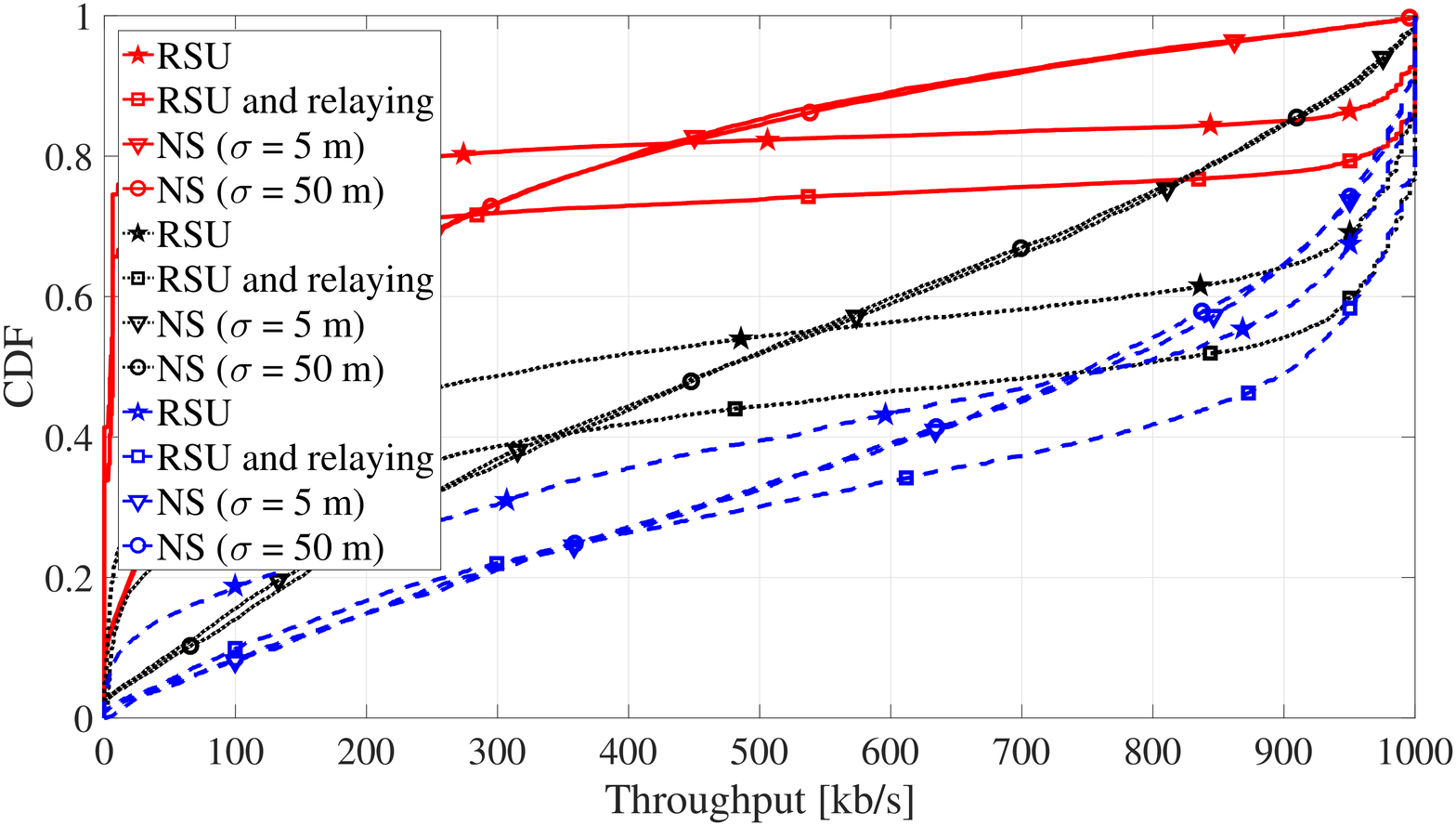}
	\caption{\label{Video}CDF of the throughput of infotainment slice for different technologies.}
\end{figure}

\begin{table}
	\centering
	\caption{PRR comparison of RSU and network slicing based V2X network for the autonomous driving slice.}%
	\label{PRR_safety}
	{
		\begin{tabular}{|p{2.3cm}|p{3.5cm}|p{1.5cm}|}
			\hline
			\textbf{Inter-vehicular distance} & \textbf{Technology} & \textbf{Probability} \\
			\hline
			& RSU  & 31.15 \%  \\ \cline{2-3}
			1-100 m& NS ($\sigma=5$ m) & 99.47 \%  \\ \cline{2-3} 
			& NS ($\sigma=50$ m) & 96.87 \%  \\ \hline
			 & RSU & 55.73 \%  \\ \cline{2-3}
			100-200 m& NS ($\sigma=5$ m) & 99.65 \%  \\ \cline{2-3}
			& NS ($\sigma=50$ m) & 99.09 \%  \\ \hline
			 & RSU & 58.14 \%  \\ \cline{2-3}
			200-300 m & NS ($\sigma=5$ m) & 99.78 \%  \\ \cline{2-3}
			& NS ($\sigma=50$ m) & 99.15 \%  \\ \hline
			
	\end{tabular}}
\end{table}

Creation of autonomous driving slice offloads the safety message traffic from the RSU to slice access point. Resources of the RSU, which were previously used to serve autonomous vehicles are now used to provide infotainment services only. Fig. \ref{Video} shows the CDF of throughput of video streaming vehicles. 
Direct RSU communication with relaying performs the best in terms of throughput because transmissions of low SINR vehicles use two high quality (V2V and V2I) links instead of one poor quality (V2I) link. On the other hand network slicing decreases the user outage probability compared to direct RSU communication.

\vspace{10pt}

\begin{figure}
	\centering
	\includegraphics[trim = 0mm 0mm 0mm 0mm, clip, width=0.5\textwidth]{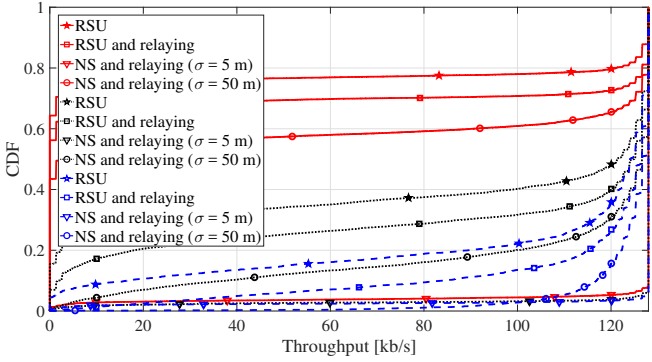}
	\caption{\label{Safety1}CDF of the throughput of autonomous  slice for different technologies.}
\end{figure}

\subsection{Combined Network Slicing and Relaying}\label{sec:CNS}
\begin{figure}
	\centering
	\includegraphics[trim = 0mm 0mm 0mm 0mm, clip, width=0.5\textwidth]{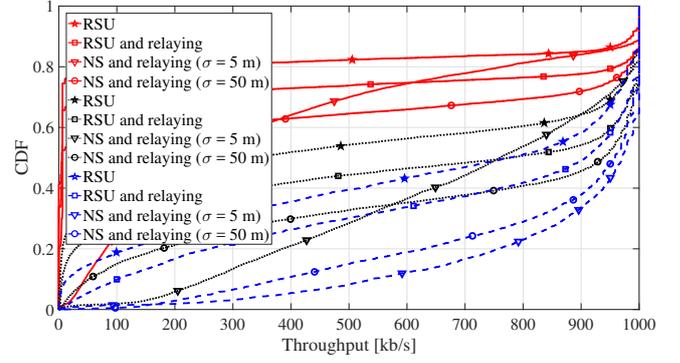}
	\caption{\label{Video1}CDF of the throughput of infotainment slice for different technologies.}
\end{figure}

This section sheds light on the performance of V2X network when both network slicing and relaying are used jointly to increase the throughput of infotainment slice. Fig. \ref{Safety1} and \ref{Video1} shows the CDF of throughput for autonomous driving and infotainment slice.  Infotainment vehicle which has high quality V2V link and which is not acting as an autonomous driving slice access point is used to relay video stream. Since autonomous driving slice access point and relaying vehicles both use the same V2V link it increases the interference observed by autonomous driving slice. 

Comparison of Fig. \ref{Safety1} and \ref{Safety} shows that the probability of a user achieving the target throughput of 128 kbps for autonomous driving slice is lowered in combined method compared to network slicing. In case of combined method 95.06 \% of vehicles achieve the target rate, while network slicing provides the target rate to 99.907 \% of the vehicles for inter-vehicular distance of 200-300 m and neighbourhood size of 5 m. Table \ref{throughput_combined} and \ref{PRR_safety} lists the PRR of different technologies in V2X network for combined method and network slicing. It is shown that an increase of interference reduces the throughput and PRR of the combined method. 

Fig. \ref{Video1} shows the CDF of infotainment slice when both network slicing and relaying are used together. While the combined approach improves the throughput of the infotainment slice it reduces the PRR of safety messages. The combined approach improves the throughput of infotainment slice by utilizing two hop high quality transmission for vehicles having low SINR. Probability of vehicles achieving the target throughput of 1000 kbps is increased by 15 \% using the combined method compared to network slicing alone for inter-vehicular distance of 200-300 m and neighbourhood size of 5 m. Similar effect of neighbourhood size can be observed from Fig. \ref{Video1} where, smaller value of sigma results in higher throughput.

\begin{table}
	\centering
	\caption{PRR comparison of RSU and combined method based V2X network for the autonomous driving slice.}%
	\label{throughput_combined}
	{
	\begin{tabular}{|p{2.3cm}|p{3.5cm}|p{1.5cm}|}
		\hline
		\textbf{Inter-vehicular distance} & \textbf{Technology} & \textbf{Probability} \\ \hline
		& RSU & 31.15 \%  \\ \cline{2-3}
		1-100 m & NS and relaying ($\sigma=5$ m) & 97.19 \%  \\ \cline{2-3} 
		& NS and relaying ($\sigma=50$ m) & 41.65 \%  \\ \hline
		& RSU & 55.73 \%  \\ \cline{2-3}
		100-200 m & NS and relaying ($\sigma=5$ m) & 98.44 \%  \\ \cline{2-3}
		& NS and relaying ($\sigma=50$ m) & 65.76 \%  \\ \hline
		& RSU & 58.14 \%  \\ \cline{2-3}
		200-300 m & NS and relaying ($\sigma=5$ m) & 98.48 \%  \\ \cline{2-3}
		& NS and relaying ($\sigma=50$ m) & 67.9 \%  \\ \hline
	\end{tabular}}
\end{table}

\section{Conclusion}\label{sec:Conclusion}
In this paper, the performance of network slicing in a vehicular network is analysed. A six lane highway layout was modelled in LTE-A system level simulator. The proposed network slicing method involves the configuration of vehicular UE functionality, to provide heterogeneous services. The network density is varied by changing the inter-vehicular distance between vehicles. Network slicing improves the performance of autonomous driving slice by many folds compared to the direct RSU communication method. When network slicing and relaying are combined, it increases the throughput of infotainment slice compared to when network slicing alone is performed. 
Since the network is interference limited relaying video streams with V2V links increases the overall interference of autonomous driving slice. Simulation result shows that both proposed network slicing approaches outperform the direct RSU communication method and attain huge improvements in term of reliability and throughput. In our future work we will investigate the effect of queuing latency on vehicular networks with network slicing.



\section*{Acknowledgment}
This research was supported by the Academy of Finland under grant 302230, High5 project number 2192/31/2016 funded by Business Finland, Bittium, Keysight, Kyynel, MediaTek, Nokia, University of Oulu and the Thule institute strategic project SAFARI.



\bibliographystyle{IEEEtran}
\footnotesize
\bibliography{Networkslicing}

\begin{thebibliography}{10}
\providecommand{\url}[1]{#1}
\csname url@samestyle\endcsname
\providecommand{\newblock}{\relax}
\providecommand{\bibinfo}[2]{#2}
\providecommand{\BIBentrySTDinterwordspacing}{\spaceskip=0pt\relax}
\providecommand{\BIBentryALTinterwordstretchfactor}{4}
\providecommand{\BIBentryALTinterwordspacing}{\spaceskip=\fontdimen2\font plus
\BIBentryALTinterwordstretchfactor\fontdimen3\font minus
  \fontdimen4\font\relax}
\providecommand{\BIBforeignlanguage}[2]{{%
\expandafter\ifx\csname l@#1\endcsname\relax
\typeout{** WARNING: IEEEtran.bst: No hyphenation pattern has been}%
\typeout{** loaded for the language `#1'. Using the pattern for}%
\typeout{** the default language instead.}%
\else
\language=\csname l@#1\endcsname
\fi
#2}}
\providecommand{\BIBdecl}{\relax}
\BIBdecl

\bibitem{ultra}
G.~Araniti, C.~Campolo, M.~Condoluci, A.~Iera, and A.~Molinaro, ``{LTE} for
  vehicular networking: a survey,'' \emph{IEEE Communications Magazine},
  vol.~51, no.~5, pp. 148--157, May 2013.

\bibitem{national}
{National Highway Traffic Safety Administration}, ``National motor vehicle
  crash causation survey,''
  \url{https://crashstats.nhtsa.dot.gov/Api/Public/ViewPublication/811059},
  report no. DOT HS 811 059, 2008.

\bibitem{V2X_IEEE}
L.~Le, A.~Festag, R.~Baldessari, and W.~Zhang, ``Vehicular wireless short-range
  communication for improving intersection safety,'' \emph{IEEE Communications
  Magazine}, vol.~47, no.~11, pp. 104--110, Nov 2009.

\bibitem{unbounded}
A.~Vinel, ``{3GPP LTE} versus {IEEE 802.11p/WAVE}: Which technology is able to
  support cooperative vehicular safety applications?'' \emph{IEEE Wireless
  Communications Letters}, vol.~1, no.~2, pp. 125--128, Apr 2012.

\bibitem{V2X_LTE2}
C.~Lottermann, M.~Botsov, P.~Fertl, and R.~Mullner, ``Performance evaluation of
  automotive off-board applications in lte deployments,'' in \emph{2012 IEEE
  Vehicular Networking Conference (VNC)}, Nov 2012, pp. 211--218.

\bibitem{speed}
{ETSI TS 122 185}, ``{LTE}; service requirements for {V2X} services ({3GPP TS
  22.185 version 14.3.0 Release 14}),''
  \url{http://www.etsi.org/deliver/etsi_en/302600_302699/30263702/01.03.00_20/en_30263702v010300a.pdf},
  Mar 2017.

\bibitem{CAM}
{ETSI EN 302 637-2}, ``Intelligent transport systems; vehicular communications;
  basic set of applications; part 2: Specification of cooperative awareness
  basic service,''
  \url{http://www.etsi.org/deliver/etsi_en/302600_302699/30263702/01.03.00_20/en_30263702v010300a.pdf},
  Mar 2017.

\bibitem{RRM}
\BIBentryALTinterwordspacing
A.~Anpalagan, M.~Bennis, and R.~Vannithamby, \emph{Design and Deployment of
  Small Cell Networks}.\hskip 1em plus 0.5em minus 0.4em\relax Cambridge
  University Press, 2015. [Online]. Available:
  \url{https://books.google.fi/books?id=ZjoACwAAQBAJ}
\BIBentrySTDinterwordspacing

\bibitem{claudia}
C.~Campolo, A.~Molinaro, A.~Iera, and F.~Menichella, ``5g network slicing for
  vehicle-to-everything services,'' \emph{IEEE Wireless Communications},
  vol.~24, no.~6, pp. 38--45, Dec 2017.

\bibitem{Petri1}
P.~Luoto, M.~Bennis, P.~Pirinen, S.~Samarakoon, K.~Horneman, and M.~Latva-aho,
  ``System level performance evaluation of {LTE-V2X} network,'' in
  \emph{European Wireless 2016; 22th European Wireless Conference}, Oulu,
  Finland, May 2016, pp. 1--5.

\bibitem{Petri2}
P.~Luoto, M.~Bennis, P.~Pirinen, S.~Samarakoon, K.~Horneman, and M.~Latva-aho, ``Vehicle clustering for improving enhanced {LTE-V2X} network
  performance,'' in \emph{2017 European Conference on Networks and
  Communications (EuCNC)}, Oulu, Finland, Jun 2017, pp. 1--5.

\bibitem{multi-unicast}
I.~Safiulin, S.~Schwarz, T.~Philosof, and M.~Rupp, ``Latency and resource
  utilization analysis for {V2X} communication over {LTE MBSFN} transmission,''
  in \emph{WSA 2016; 20th International ITG Workshop on Smart Antennas},
  Munich, Germany, March 2016, pp. 1--6.

\bibitem{RRM2-Ikram}
M.~I. Ashraf, C.-F. Liu, M.~Bennis, and W.~Saad, ``Towards low-latency and
  ultra-reliable vehicle-to-vehicle communication,'' in \emph{2017 European
  Conference on Networks and Communications (EuCNC)}, Oulu, Finland, Jun 2017,
  pp. 1--5.

\bibitem{slice1}
P.~Rost, C.~Mannweiler, D.~S. Michalopoulos, C.~Sartori, V.~Sciancalepore,
  N.~Sastry, O.~Holland, S.~Tayade, B.~Han, D.~Bega, D.~Aziz, and H.~Bakker,
  ``Network slicing to enable scalability and flexibility in 5{G} mobile
  networks,'' \emph{IEEE Communications Magazine}, vol.~55, no.~5, pp. 72--79,
  May 2017.

\bibitem{metis_slicing}
\BIBentryALTinterwordspacing
``{METIS} deliverable {D}7.3 {F}inal 5{G} visualization,'' Jun 2017. [Online].
  Available:
  \url{https://metis-ii.5g-ppp.eu/wp-content/uploads/deliverables/METIS-II_D7.3_V1.0.pdf}
\BIBentrySTDinterwordspacing

\bibitem{WINII}
\BIBentryALTinterwordspacing
``{WINNER II channel models, D1.1.2 V1.0}.'' [Online]. Available:
  \url{http://www.cept.org/files/1050/documents/winner2-finalreport.pdf}
\BIBentrySTDinterwordspacing

\bibitem{V2V_pathloss}
J.~Karedal, N.~Czink, A.~Paier, F.~Tufvesson, and A.~F. Molisch, ``Path loss
  modeling for vehicle-to-vehicle communications,'' vol.~60, no.~1, pp.
  323--328, Jan 2011.

\bibitem{clustering_tutorial}
\BIBentryALTinterwordspacing
U.~von Luxburg, ``A tutorial on spectral clustering,'' \emph{CoRR}, vol.
  abs/0711.0189, 2007. [Online]. Available:
  \url{http://arxiv.org/abs/0711.0189}
\BIBentrySTDinterwordspacing

\bibitem{Gaussian_similarity}
A.~L. Yuille and N.~M. Grzywacz, ``The motion coherence theory,'' in
  \emph{Second International Conference on Computer Vision,}, Dec 1988, pp.
  344--353.

\bibitem{3GPP_frequency}
\BIBentryALTinterwordspacing
``{3GPP TR 36.785 V14.0.0}. {Vehicle to Vehicle {(V2V)} services based on LTE
  sidelink; User Equipment {(UE)} radio transmission and reception }.''
  [Online]. Available: \url{http://www.3gpp.org/dynareport/36785.htm}
\BIBentrySTDinterwordspacing

\end{thebibliography}

%



\end{document}